\documentclass[useAMS,usenatbib]{mn2e}
\usepackage{amsmath}
\usepackage{blindtext}
\usepackage{graphicx}
\usepackage{amssymb}
\usepackage{multirow}
\usepackage{verbatim}   
\usepackage{color}     
\usepackage{subfigure} 
\usepackage{hyperref}
\usepackage{latexsym}
\usepackage{caption}
\usepackage{wrapfig}
\usepackage{soul}
\usepackage{natbib}
\usepackage{float} 
\usepackage{placeins}

\title[Beta Function Quintessence Cosmological Parameters and Fundamental Constants II: 
Exponential and Logarithmic Dark Energy Potentials]{Beta Function Quintessence Cosmological Parameters and Fundamental Constants II: Exponential and Logarithmic Dark Energy Potentials}
\author[Rodger I. Thompson]{Rodger I. Thompson$^{1}$\thanks{E-mail:
rit@email.arizona.edu (RIT)}\\
$^{1}$Steward Observatory, University of Arizona, Tucson, AZ 85721, USA}

\begin{document}

\date{Accepted xxxx. Received xxxx; in original form xxxx}

\pagerange{\pageref{firstpage}--\pageref{lastpage}} \pubyear{2016}

\maketitle

\label{firstpage}

\begin{abstract}
This paper uses the beta function formalism to extend the analysis of quintessence
cosmological parameters to the logarithmic and exponential dark energy potentials.
The previous paper \citep{thm18} demonstrated the formalism using power and
inverse power potentials.  The essentially identical evolution of the Hubble parameter for
all of the quintessence cases and $\Lambda$CDM is attributed to the flatness of the
quintessence dark energy potentials in the dark energy dominated era. The Hubble parameter
is therefore incapable of discriminating between static and dynamic dark energy.  Unlike 
the other three potentials considered in the two papers the logarithmic dark energy
potential requires a numerical integration in the formula for the superpotential rather 
than being an analytic function.  The dark energy equation
of state and the fundamental constants continue to be good discriminators between static
and dynamical dark energy.  A new analysis of quintessence with all four of the potentials 
relative the swampland conjectures indicates that the conjecture on the change in the
scalar field is satisfied but that the conjecture on the change of the potential is not.
\end{abstract}


\maketitle

\section{Introduction} \label{s-intro} 
This is the second of two papers using the beta function methodology to produce accurate 
analytic solutions from model dark energy potentials in a quintessence cosmology.   The
first paper \citep{thm18}, hereinafter paper I, examined solutions for power and inverse power law 
potentials. This work extends the analysis to logarithmic and exponential potentials. The 
analytic nature of the solutions provides the means to calculate solutions for other
values of the input parameters such as $H_0$ and $\Omega_{m_0}$ in a flat universe
for comparison with observations.

Exact analytic solutions for specific dark energy potentials are often mathematically intractable
\citet{nar17} but the beta function formalism  \citep{bin15, cic17} provides a method for achieving 
accurate analytic solutions using beta potentials $V_b(\phi)$ that are accurate, but not exact, 
representations of model potentials $V_m(\phi)$.  In many cases numerical calculations can 
provide solutions for specific cases.  Such solutions, however, often do not readily reveal the 
basic physics in play nor do they provide easily calculable solutions for alternative input parameters.  
The particular potentials examined here are the logarithmic
\begin{equation} \label{eq-vl}
V_m(\phi) \propto (\frac{\ln(\phi)}{\ln(\phi_0)})^{\beta_l}
\end{equation}
and exponential 
\begin{equation} \label{eq-ve}
V_m(\phi) \propto \exp{[-\beta_e(\phi-\phi_0)]}
\end{equation}
potentials where $\beta_l$ and $\beta_e$ are real, positive constants.

The methodology follows the descriptions in \citet{bin15,cic17},
particularly \citet{cic17} who explicitly include matter as well as dark energy.  The details
of the analysis are given in paper I and will not be repeated here except for clarity.
This work concentrates on the "late time" evolution of the universe which is taken to be the 
time between a scale factor of 0.1 and 1.0 corresponding to redshifts between zero and nine.  
A flat universe is assumed with $H_0 = 70$  km/sec per megaparsec. The current ratio of 
the dark energy density to the critical density $\Omega_{\phi_0}$ is set to 0.7 where $\phi_0$ 
is the current value of the scalar $\phi$.   The current values of the dark energy equation of 
state are set to $w_0=(-0.98, -0.96, -0.94, -0.92,-0.90)$ as was done in paper I.  The last two 
values of $w_0$ are unlikely but are included to determine the limits on the validity of the 
solutions.  In the exponential model potential the value of $w_0$ determines the value of 
$\beta_e$ removing one degree of freedom.  In paper I $\kappa=\frac{\sqrt{8\pi}}{m_{pl}}$
was set to one, however, in this paper natural units are used with $m_{pl}$, the Planck mass,
set to one.  This makes the units of the scalar $\phi$ the planck mass rather than $1/\kappa$.
A section on where the quintessence cases considered here and in paper I dwell relative to the
swampland conjectures has also been added.

\section{Quintessence} \label{s-q}
Quintessence is characterized by an action of the form
\begin{equation} \label{eq-act}
S=\int d^4x \sqrt{-g}[\frac{R}{2}-\frac{1}{2}g^{\mu\nu}\partial_{\mu}\partial_{\nu}\phi -V(\phi)]
+S_m
\end{equation}
where $R$ is the Ricci scalar, $g$ is the determinant of the metric $g^{\mu\nu}$, $V(\phi)$ 
is the dark energy potential, and, $S_m$ is the action of the matter fluid.  Different types
of quintessence are defined by different forms of the dark energy potential.   
The quintessence dark energy density, $\rho_{\phi}$, and pressure, $p_{\phi}$, are  given
by
\begin{equation} \label{eq-rhop}
\rho_{\phi} \equiv \frac{\dot{\phi}^2}{2}+V(\phi), \hspace{1cm} p_{\phi}  \equiv \frac{\dot{\phi}^2}{2}-V(\phi)
\end{equation}

\section{The Beta Function} \label{s-beta}
The beta function is defined as the derivative of the scalar $\phi$ with respect to the natural log
of the scale factor $a$ \citep{bin15}
\begin{equation} \label{eq-beta}
\beta(\phi) \equiv \frac{\kappa d \phi}{d \ln(a)} =\kappa \phi'
\end{equation}
where $\kappa=\frac{\sqrt{8\pi}}{m_{pl}}$ and  the prime on the right hand term 
denotes the derivative with respect to the natural log of the scale factor except when 
it denotes the integration variable inside an integral as in equation~\ref{eq-sfi}. As
noted in the introduction paper I set $\kappa$ to one as is often done in the cosmological
literature.  Here instead the Planck mass is set to one leading to the scalar $\phi$ being
expressed in units of the Planck mass, a difference of $\sqrt{8\pi} \approx 5$ from paper
I.  In the following $k$ is used to denote $\sqrt{8\pi}$ in an equation.  Note that $\phi$
now has the dimensions of $m_{pl}$ and that $\kappa \phi$ is dimensionless. 

The dark energy equation of state $w=\frac{p_{\phi}}{\rho_{\phi}}$ for
quintessence is given by \citet{nun04} 
\begin{equation} \label{eq-nun4}
w+1 =\frac{k^2\phi'^2}{3 \Omega_{\phi}} =\frac{k^2 \beta^2(\phi)}{3 \Omega_{\phi}}
\end{equation}
For the logarithmic potentials this equation provides the boundary condition to determine
the current value of the scalar $\phi_0$.  For the exponential potential eqn.~\ref{eq-nun4}
determines $\beta_e$ as discussed in section~\ref{s-es}.

The beta function is not an arbitrary function of $\phi$ and $a$ but is determined by the 
model dark energy potential $V_m(\phi)$ such that \citep{cic17}
\begin{equation} \label{eq-bv}
V_m(\phi)=\exp\{-\int k\beta(\phi)dk\phi\}
\end{equation}

\subsection{Beta Functions from the Potentials} \label{ss-bpot}
From eqn.~\ref{eq-bv} the appropriate beta function is the logarithmic derivative of the
potential.   Using the potentials listed in the introduction the logarithmic beta function
is
\begin{equation} \label{eq-lb}
\beta(\phi) = (\frac{-\beta_l}{k \phi \ln(k\phi)})
\end{equation}
The exponential beta function is simply
\begin{equation} \label{eq-ilb}
\beta(\phi) =\frac{ \beta_e}{k}
\end{equation}
Five $\beta_l$ values are considered, the integers one through five.  The $\beta_e$ values 
are set by the five values of $w_0$

\section{Evolution of the Scalar} \label{s-es}
An important feature of the beta function formalism is that the specification of the beta function,
along with a boundary condition determines the evolution of the scalar with respect to the scale
factor $\phi(a)$.  

\subsection{The Scalar as a Function of the Scale Factor (Logarithmic)} \label{ss-sfl}
The beta function, eqn.~\ref{eq-beta}, provides the differential equation for $\phi$ as a
function of the scalar $a$.  For the logarithmic potential
\begin{equation} \label{eq-ld}
k^2 \phi \ln(k\phi) d \phi =  -\beta_l d\ln(a)
\end{equation}
Integrating both sides
\begin{equation} \label{eq-sfi}
\int_{\phi_0}^{\phi}k^2 \phi'\ln(k\phi') d\phi' =-\beta_l\int_1^a d\ln(a')
\end{equation}
gives
\begin{equation} \label{eq-il}
\frac{k^2 \phi^2}{2}(\ln(k\phi)-\frac{1}{2})=-\beta_l\ln(a)+\frac{k^2\phi_0^2}{2}(\ln(k\phi_0)-\frac{1}{2})
\end{equation}
where $\phi_0$ is the current value of the scalar.  Denoting the 
right hand term of the equation by $Q$ the scalar is given by
\begin{equation} \label{eq-pl}
k\phi=\pm\sqrt{\frac{2Q}{PL(\frac{2Q}{e})}}
\end{equation}
The term $PL$ in eqn.~\ref{eq-pl} stands for the Product Log, more commonly known as the
Lambert $W(x)$ function, the solution to $We^W=x$.  Here the ProductLog term, 
used by Mathematica, is retained to avoid confusion with the superpotential $W(\phi)$ 
introduced later.
The value of $\phi_0$ is determined by the current value of the dark energy equation of
state $w_0$ using eqn.~\ref{eq-nun4}
\begin{equation} \label{eq-lpo}
k\phi_0\ln(k\phi_0)=\frac{\pm\beta_l}{\sqrt{3\Omega_{\phi_0}(w_0+1)}}
\end{equation}
where $\Omega_{\phi_0}$ is the current ratio of the dark energy density to the critical
density.  The solution to eqn.~\ref{eq-lpo} again uses the PL function
\begin{equation} \label{eq-plpo}
k\phi_0=\frac{\frac{\pm\beta_l}{\sqrt{3\Omega_{\phi_0}(w_0+1)}}}{PL(\frac{\pm \beta_l}{\sqrt{3\Omega_{\phi_0}(w_0+1)}})}
\end{equation}
The Product Log does not have positive real solutions for negative arguments.  The definition
of the logarithmic beta function assumes that $\beta_l$ is a positive real number, therefore,
the positive square root is chosen in eqns.~\ref{eq-pl},~\ref{eq-lpo} and~\ref{eq-plpo}.
None of the three equations accommodate phantom solutions where $(w+1)<0$.

\begin{figure}
\scalebox{.6}{\includegraphics{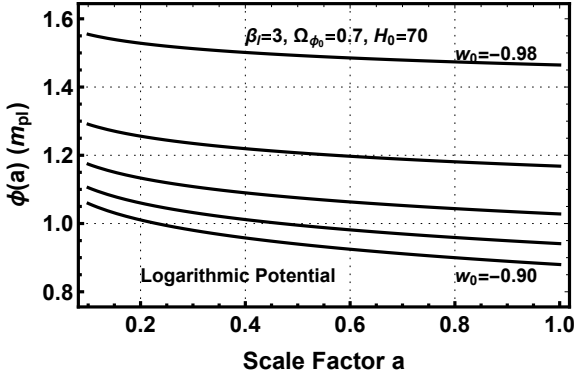}}
\caption{The evolution of the scalar field $\phi$ as a function of the scalar $a$ for the 
logarithmic beta function with $\beta_l = 3.0$ for the five values of $w_0$ listed in the 
introduction.}
\label{fig-lib}
\end{figure}

\begin{figure}
\scalebox{.6}{\includegraphics{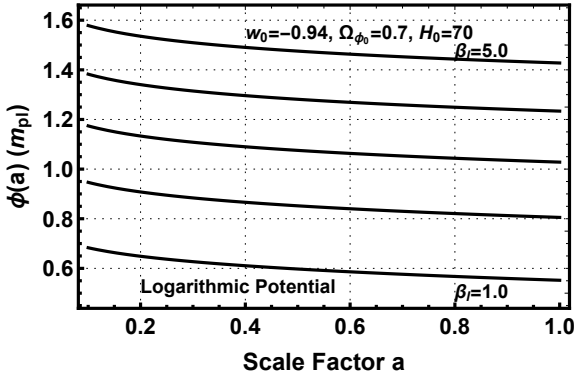}}
\caption{The evolution of the scalar field $\phi$ as a function of the scalar $a$ for the  
logarithmic beta function with the five values of $\beta_l$ and $w_0=-0.94$.}
\label{fig-liw}
\end{figure}

Figures~\ref{fig-lib} and~\ref{fig-liw} show the evolution of the scalar $\phi$ for the  
logarithmic beta function with $\beta_l$ held constant at 3 in fig.~\ref{fig-lib} for
the five values of $w_0$ and $w_0$ is held constant at -0.94 in fig.~\ref{fig-liw}
for the five values of $\beta_l$.
Even though $\phi_0$ changes significantly with the value of $\beta_l$, the scalar $\phi$  
evolves relatively little over $a$ between 0.1 and 1.    

\subsection{The Scalar as a Function of the Scale Factor (Exponential)} \label{ss-sfe}
\begin{figure}
\scalebox{.6}{\includegraphics{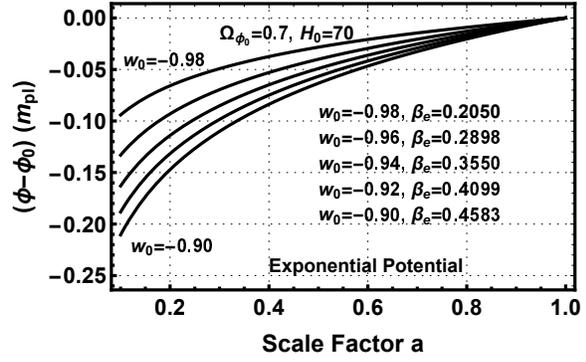}}
\caption{The evolution of the scalar field $(\phi - \phi_0)$ as a function of the scalar $a$ 
for the exponential beta function for the five values of $w_0$ listed in the introduction.}
\label{fig-exphi}
\end{figure}

The exponential potential, $V(\phi) \propto \exp[\beta_e(\phi-\phi_0)]$ is the dark
energy potential for slow roll quintessence when the first slow roll parameter,
$\frac{1}{V}\frac{dV}{d \phi}$ is held constant eg. \citet{sch08}.  The beta function
for the exponential potential, $\beta_e$, is unique in that it is a constant and not a
function of $\phi$.  Unlike all of the previous cases $\beta_e$ can not be set arbitrarily.
The value of $\beta_e$ is set by eqn.~\ref{eq-nun4} 
\begin{equation} \label{eq-bew}
\beta_e =\sqrt{3 \Omega_{\phi_0} (w_0+1)}
\end{equation}
independent of $\phi$ or $\phi_0$, therefore there is no boundary condition to set
$\phi_0$.  The solutions for the relevant cosmological parameters and fundamental
constants are all functions of $(\phi - \phi_0)$ therefore it is the appropriate parameter
rather than the absolute values of $\phi$ and $\phi_0$.  From the exponential potential
beta function
\begin{equation} \label{eq-exppo}
k(\phi - \phi_0) = \beta_e \ln(a)
\end{equation}
The evolution of $(\phi - \phi_0)$ is shown in fig.~\ref{fig-exphi}.
The values of $\beta_e$ for the appropriate values of $w_0$ are listed in fig.~\ref{fig-exphi}
and are all less than one.

An anonymous referee has pointed out that a constant beta function never reaches a
fixed de Sitter point which requires a beta function value of zero.  The referee
also mentioned that for a small value of the beta function, as is found here, that space time
is evolving toward a power law geometry that might have interesting consequences in 
holography as discussed in \citet{cic18}.

\section{The Evolution of the Beta Function} \label{s-ebeta}
\begin{figure}
\scalebox{.6}{\includegraphics{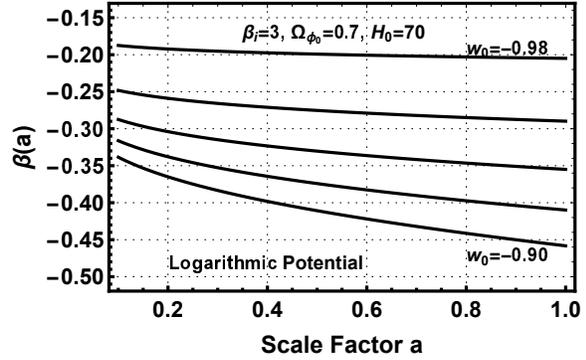}}
\caption{The evolution of $\beta(a)$ as a function of the scalar $a$ for the 
logarithmic potential with $\beta_l=3$ for the five values of $w_0$ listed in the introduction.}
\label{fig-lgbetab3}
\end{figure}

\begin{figure}
\scalebox{.6}{\includegraphics{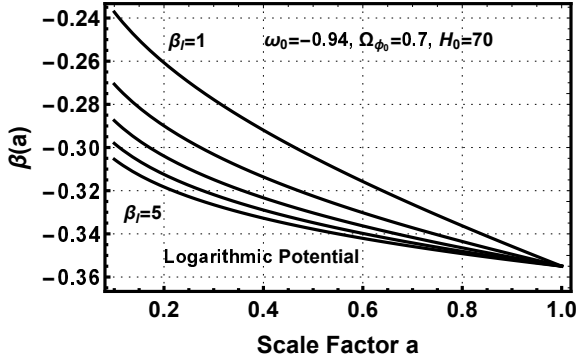}}
\caption{The evolution of $\beta(a)$ as a function of the scalar $a$ for the 
logarithmic potential with $w_0=-0.94$ for the five values of $\beta_l$ listed in the introduction.}
\label{fig-lgbetab}
\end{figure}

In the beta function formalism many of the cosmological parameters depend on the form
of the beta function.  Figures~\ref{fig-lgbetab3} and~\ref{fig-lgbetab} 
display the evolution of the logarithmic potential beta functions for the five values of $w_0$
with $\beta_l=3$, fig.~\ref{fig-lgbetab3} and for the five values of $\beta_i$ with $w_0=-0.94$,
fig.~\ref{fig-lgbetab}.  The logarithmic beta functions are negative and between -0.1 and -0.5
for scale factors between 0.1 and 1.

\begin{figure}
\scalebox{.6}{\includegraphics{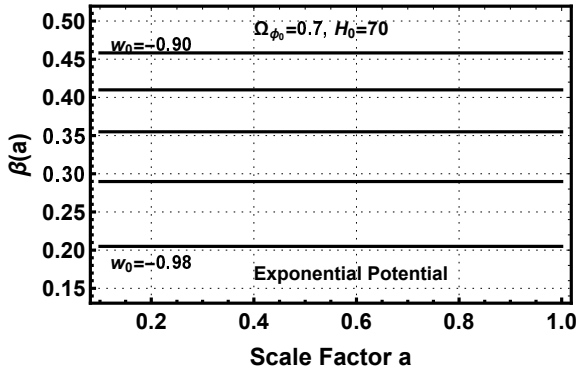}}
\caption{The evolution of $\beta(a)$ as a function of the scalar $a$ for the exponential
potential.}
\label{fig-exbeta}
\end{figure}

Figure~\ref{fig-exbeta} shows the evolution of the exponential potential beta function for
the five $\beta_e$, $w_0$ pairs.  The values are positive and constant which simplifies 
several of the subsequent calculations.

\section{The Potentials} \label{s-pot}
In the beta function formalism two different types of potentials play a prominent role.  
The first is the dark energy potential in the action $V(\phi)$ that does not depend on 
matter.  The second, in analogy with particle physics, is termed the superpotential $W$ 
given by
\begin{equation} \label{eq-W}
W(\phi) = -2H(\phi) = -2\frac{\dot{a}}{a}
\end{equation}
Even though the Hubble parameter $H$ is the parameter of interest $W$ is utilized 
here to be consistent with the literature on beta functions. Both the dark energy potential 
$V(\phi)$ and the superpotential $W(\phi)$ can be expressed in terms of $\beta(\phi)$ 
\citep{cic17} by
\begin{equation} \label{eq-wphi}
W(\phi) = W_0 \exp\{-\frac{1}{2}\int_{\phi_0}^{\phi}\beta(k\phi')kd\phi'\}
\end{equation}
and
\begin{equation} \label{eq-v}
V(\phi) = \frac{3}{4 k^2} W_0^2 \exp\{-\int_{\phi_0}^{\phi}\beta(k\phi')kd\phi'\}(1-\frac{\beta^2(k\phi)}{6})
\end{equation}
where $W_0$ is the current value of $W$ equal to $-2H_0$.  Note that the superpotential is
always denoted as a capital $W$ and the dark energy equation of state by a lower case $w$.
The potential in eqn.~\ref{eq-v} is referred to as the beta potential of the model potential.  It
differs from the model potential by the factor of $(1-\frac{\beta^2(\phi)}{6})$.  As long as this
factor is close to one the beta potential is an accurate, but not exact, representation of the model
potential.

\subsection{The Logarithmic Potential} \label{ss-lp}
\begin{figure}
\scalebox{.6}{\includegraphics{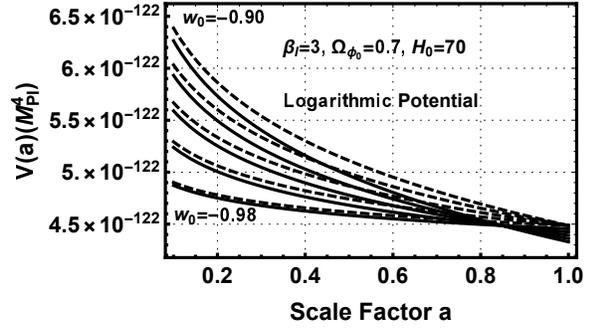}}
\caption{The evolution of the model logarithmic potential with $\beta_l =3$ is shown by 
the dashed lines and the solid lines indicate the evolution of the beta logarithmic potential.}
\label{fig-vl}
\end{figure}

The model logarithmic potential is given by
\begin{equation} \label{eq-mlp}
V_m(\phi)=\frac{3}{4 k^2}W_0^2(\frac{\ln(k\phi)}{\ln(k\phi_0)})^{\beta_l}
\end{equation}
with the beta function shown in eqn.~\ref{eq-lb}. The logarithmic beta potential is given by
\begin{equation} \label{eq-blp}
V_b(\phi)=\frac{3}{4 k^2}W_0^2(\frac{\ln(k\phi)}{\ln(k\phi_0)})^{\beta_l}(1- \frac{\beta_l^2}
{6(k\phi \ln(k\phi))^2})
\end{equation}
The logarithmic potential is decreasing as the scale factor increases.  Figure~\ref{fig-vl} shows 
the potential with $\beta_l$ fixed at 3 for the five different values of $w_0$.  The solid lines in 
fig.~\ref{fig-vl} show the beta potential which follows the model potential (dashed) quite 
well, particularly for values of $w_0$ close to minus one.  The accuracy of the fit is quantified 
in section~\ref{ss-fit} for all of the potentials.

\subsection{The Exponential Potential} \label{ss-exp}
The model potential is of the form
\begin{equation} \label{eq-exppot}
V_m(\phi)=\frac{3}{4 k^2}W_0^2 \exp(-\beta_ek(\phi-\phi_0))
\end{equation}
with a beta potential of
\begin{equation} \label{eq-expbetapot}
V_b(\phi)=\frac{3}{4 k^2}W_0^2 \exp(-\beta_ek(\phi-\phi_0))(1-\frac{\beta_e^2}{6})
\end{equation}
\begin{figure}
\scalebox{.6}{\includegraphics{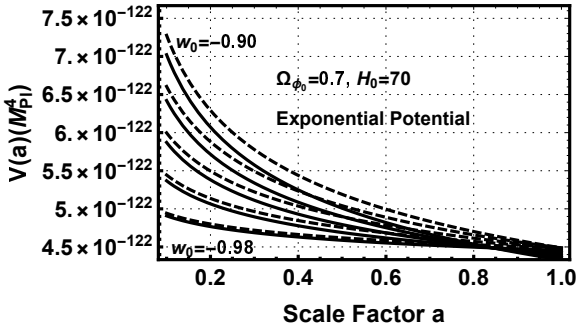}}
\caption{The evolution of the model exponential potential is shown by the dashed lines 
and the solid lines indicate the evolution of the beta exponential potential.}
\label{fig-vexp}
\end{figure}
Figure~\ref{fig-vexp} shows the evolution of exponential model and beta potentials.

\subsection{Normalization} \label{ss-norm}
It is clear that the beta dark energy potentials have the desired model potentials 
multiplied by $(1-\frac{\beta(\phi)^2}{6})$ which produces both an offset 
and a deviation from the model potentials.  The deviation is expected to be small since 
$\frac{\beta(\phi)^2}{6}$ is much less than one in most cases.  In paper I the potential 
was normalized to be $\frac{3}{4}W_0^2$ at a scale factor
of one producing a potential slightly different than the true beta potential.  In this work
that practice has been abandoned and no normalization has been applied.  As a result
the beta potentials shown in figs.~\ref{fig-vl}, and~\ref{fig-vexp} cross over each 
other at $a \approx 0.8$ due to the $\frac{\beta(\phi)^2}{6}$ term.

\subsection{Accuracy of Fit} \label{ss-fit}
The cosmological parameters derived by the beta function formalism are only useful if the
beta potentials accurately represent the model potentials.  Figures~\ref{fig-lgferr} 
and~\ref{fig-expferr} show the fractional deviation of the beta potentials  from the model 
potentials to quantify the deviations of the beta potentials from the model potentials.  For
the logarithmic potential the minimum, median and maximum $\beta_l$ values are shown 
with $w_0$ values equal to -0.98, -0.94 and -0.9 to show the extremes without excessive 
overlap of tracks in the figures. For the exponential potential all of the cases are shown since
they do not overlap  In paper I a conservative limit of only accepting solutions with fractional 
deviations of $1\%$ or less was adopted.  In this paper that limit is expanded to $4\%$ which 
is a higher accuracy than the accuracy of most of the available observational data.

\subsubsection{The Logarithmic Beta Potential Fractional Error} \label{sss-lfe}
\begin{figure}
\scalebox{.6}{\includegraphics{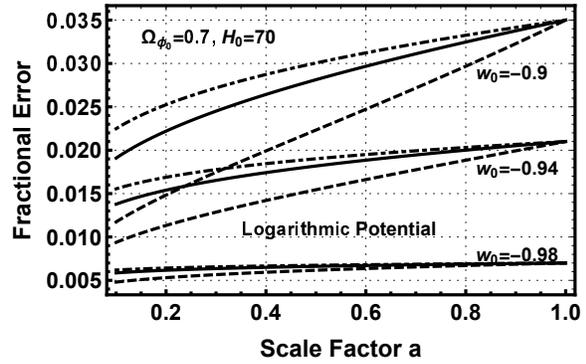}}
\caption{The fractional deviation of the beta logarithmic law potentials from the model potentials 
with $\beta_l=1$, dashed lines, $\beta_l =3.0$, solid lines, and $\beta_p =5.0$, dot 
dashed lines. For each $\beta_l$ the tracks are marked with the value $w_0$ at the
end.}
\label{fig-lgferr}
\end{figure}
The primary feature of the logarithmic potential fractional deviation in fig.~\ref{fig-lgferr}  
is that all of the cases are within the acceptable error of 0.04.
Unlike the normalized cases of paper I the highest fractional deviation for the logarithmic
beta potential is at a scale factor of one increasing for values of $w_0$ further from minus
one but independent of the value of $\beta_l$.  The evolution away from $a=1$ is dependent
on $\beta_l$ but is decreasing for lower values of $a$.  All of the logarithmic cases are
therefore retained in the subsequent analysis.

\subsubsection{The Exponential Beta Potential Fractional Errors} \label{sss-efe}
\begin{figure}
\scalebox{.6}{\includegraphics{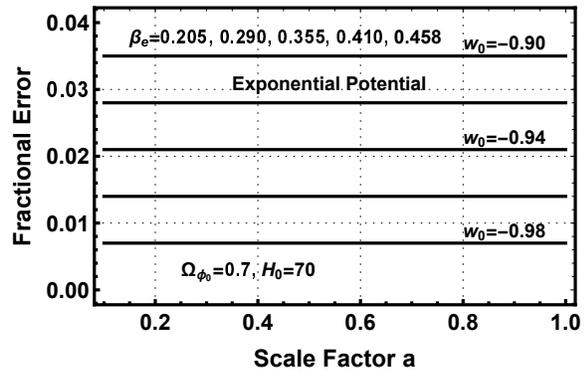}}
\caption{The same as for fig.~\ref{fig-lgferr} except for the exponential law potentials.
All five values of $\beta_e$ are shown with the values of $w_0$ marked on the figure.}
\label{fig-expferr}
\end{figure}

The exponential beta potential fractional errors shown in fig.~\ref{fig-expferr} are set by 
the values of $w_0$ which also sets the value of $\beta_e$.
As expected the fractional deviations of the exponential beta potential are independent of
the scale factor since $\beta(\phi)$ is constant for a given $w_0$ and all fall in the acceptable
range.  As with the logarithmic beta potential all of the exponential cases are retained in
the subsequent analysis.

\section{The Matter Density} \label{s-rhom} 
The dark energy potentials are independent of matter but
both baryonic and dark matter must be taken into account to calculate accurate analytic solutions 
for fundamental constants and cosmological parameters. Matter is represented by the $S_m$ 
term in the action, eqn.~\ref{eq-act}. From \citet{cic17} and paper I the matter density as a 
function of the scalar is given by
\begin{equation} \label{eq-rhomphi}
\rho_m(\phi)=\rho_{m_0}\exp(-3\int_{\phi_0}^{\phi} \frac{d \phi'}{\beta(\phi')})
\end{equation}
where $\rho_{m_0}$ is the present day mass density.
Different beta functions produce different functions for $\rho_m$ as a function of $\phi$ hiding
the universality of the matter density when expressed as a function of the scale factor $a$
\begin{equation} \label{eq-rhoma}
\rho_m(a)=\rho_{m_0}\exp(-3\int_1^a d\ln(a') )= \rho_{m0}a^{-3}
\end{equation}
as expected, independent of $\beta(\phi)$.

\section{The Superpotential $W$ and the Hubble Parameter $H$} \label{s-wh}
From eqn.~\ref{eq-W} it is obvious that calculating the superpotential $W$ is equivalent
to calculating the Hubble Parameter $H$. As shown in \citet{cic17} and paper I the 
differential equation for $W$ with matter is
\begin{equation} \label{eq-difw}
WW_{,\phi} + \frac{1}{2} \beta W^2 = -2\frac{\rho_m}{\beta}
\end{equation}
where the notation $_{,\phi}$ indicates the derivative with respect to the scalar $\phi$. 
Paper I includes two specific examples, the power and inverse power law potentials and
their related beta functions.  Here a more general solution is presented that gives a better
insight of the process.  The solutions to eqn.~\ref{eq-difw} utilize integrating factors
$f(x)$ where $x=k\phi$ for ease of notation.  The integrating factors multiply both
sides of eqn.~\ref{eq-difw} to create an exact equation that can be integrated.
The exact form on the left of the equation has the form of the left side of
eqn.~\ref{eq-lif}.  The right side is then integrated to provide the solution for W.
\begin{equation} \label{eq-lif}
\frac{d}{dx}(\frac{1}{2}W^2(x) f(x)) = -2f(x)\frac{\rho_m(x)}{\beta(x)}
\end{equation}
Comparison with eqn.~\ref{eq-difw} shows that the integrating factor must satisfy
\begin{equation} \label{eq-ff}
\frac{d f(x)}{dx} = \beta(x) f(x)
\end{equation}
which determines $f(x)$.  Writing the equation out as the equality of two differentials
gives
\begin{equation} \label{eq-fdx}
d(W^2(x)f(x)) = -4f(x)\frac{\rho_m(x)}{\beta(x)}dx
\end{equation}
Integrating both sides of eqn.~\ref{eq-fdx} gives
\begin{equation} \label{eq-ww}
W^2(x) f(x) - W_0^2 f(x_0) = -4\int_{x_0}^xf(x)\frac{\rho_m(x)}{\beta(x)}dx
\end{equation}

Equation~\ref{eq-ww} can be solved as a function of $x$ or the much more useful
function of $a$ using $x(a)$ from eqns.~\ref{eq-plpo} and~\ref{eq-exppo} and the
much simpler $\rho_m(a)$ from eqn.~\ref{eq-rhoma}. The beta function provides
the conversion of $dx$ on the right hand side of eqn.~\ref{eq-fdx} to $da$
\begin{equation} \label{eq-dxa}
dx = \beta(x(a))d\ln(a) = \beta(x(a))\frac{da}{a}
\end{equation}
The beta function in eqn.~\ref{eq-dxa} cancels the beta function on the right hand side
of eqn.~\ref{eq-fdx}. The right hand side of eqn.~\ref{eq-fdx} is now a function of $a$
rather than $x$ and the integral of the right hand side is
\begin{equation} \label{eq-irh}
-4\rho_{m_0}\int_1^a f(x(a'))a'^{-4}da'
\end{equation}
After integrating the left side of eqn~\ref{eq-fdx} and rearranging the final answer is
\begin{equation} \label{eq-waf}
W(a)=-\{-4\frac{\rho{m_0}}{f(x(a))}\int_1^a f(x(a'))a'^{-4}da +W^2_0\frac{f(x(a=1)}{f(x(a))}\}^{\frac{1}{2}}
\end{equation}

The integrating factors for the logarithmic and exponential potentials are
\begin{align} \label{eq-if}
(\ln(k \phi))^{-\beta_l} \hspace{1cm} logarithmic \nonumber \\
\exp[\beta_e k (\phi - \phi_0)] \hspace{1cm} exponential 
\end{align}
The integral in eqn.~\ref{eq-irh} for the exponential integrating factor is quite simple and analytic.
The integral for the logarithmic integrating factor is not analytic and must be done numerically 
since it contains the PL function for $x(a)$ given in eqn.~\ref{eq-pl}. 
\begin{align} \label{eq-wa}
W(a)=-[-4\rho_{m_0}(\ln(k\phi(a))^{\beta_l}\int^a_1(\ln(k\phi(a'))^{-\beta_l}a'^{-4}da' \nonumber \\
+W_0^2(\frac{\ln(k\phi(a))}{\ln(k\phi_0)})^{\beta_l}]^{\frac{1}{2}} \hspace{0.2cm} log  \nonumber \\
W(a)=-[\frac{-4 \rho_{m_0}}{\beta_e^2 -3}(a^{-3}-a^{-\beta_e^2}) +W_0^2a^{-\beta_e^2}]^{\frac{1}{2}} \hspace{0.2cm}exp
\end{align}
where $k\phi(a)$ is given by eqn.~\ref{eq-pl} for the logarithmic potentials.
The superpotential is a negative quantity therefore the negative solution of the 
square roots in eqns.~\ref{eq-wa} are used.

\subsection{The Hubble Parameter as a Function of the Scale Factor} \label{ss-eh}
\begin{figure}
\scalebox{.6}{\includegraphics{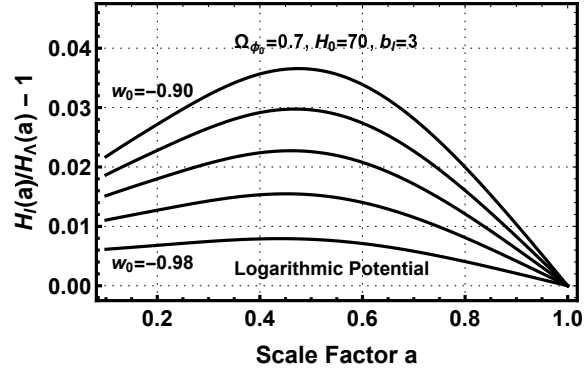}}
\caption{The ratio of the logarithmic potential evolution of the Hubble parameter $H_l(a)$.
$\beta_l$ is held constant at 3 and all five of the $w_0$ values are plotted.}
\label{fig-lh}
\end{figure}

\begin{figure}
\scalebox{.6}{\includegraphics{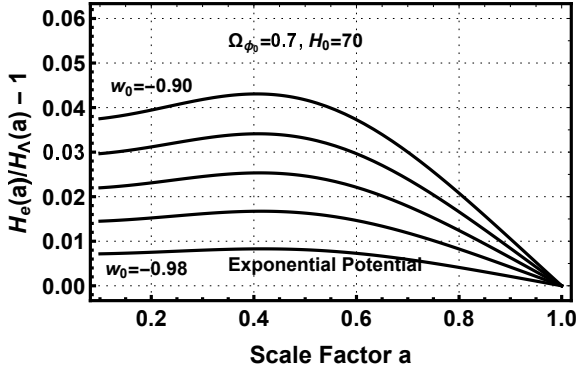}}
\caption{The same as in fig.~\ref{fig-lh} except for the five exponential potential cases.}
\label{fig-le}
\end{figure}

The Hubble parameter is simply $-\frac{W(a)}{2}$.  As was found in paper I for the
power and inverse power law potentials the evolution of the Hubble parameter for
the logarithmic and exponential potentials is indistinguishable from the $\Lambda$CDM 
evolution at the scale of the plots.  To highlight the true differences figure~\ref{fig-lh} 
shows the ratio of the Hubble parameter for logarithmic potential to the $\Lambda$CDM
minus one as a function of the scale factor. In fig.~\ref{fig-lh} $\beta_l$ is held constant
at three and each of the five values of $w_0$ are plotted
The same ratio is plotted for the five exponential potential cases in fig.~\ref{fig-le}.
In both the logarithmic and exponential cases the deviation from the $\Lambda$CDM case is
small and peaks at $a \approx 0.5$ as expected.  The similarity of the Hubble parameter
evolution for a dynamic quintessence cosmology to the static $\Lambda$CDM cosmology
makes it a poor discriminator between the two cases.  One percent accuracy observations
of $H(a)$ at redshifts near one are required to distinguish between the two.  The reason
for the similarity of the evolutions is given in the next section.

\subsection{The Evolution of the Dark Energy Density} \label{ss-ded}
\begin{figure}
\scalebox{.6}{\includegraphics{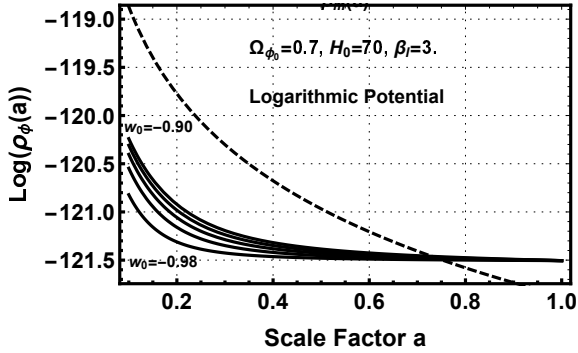}} 
\caption{The $log_{10}$ of the dark energy density values as a function of the scale factor
for the logarithmic potential.  The dashed line is the matter density which decreases below 
the dark energy density near a scale factor of 0.75}
\label{fig-lgrde}
\end{figure}

\begin{figure}
\scalebox{.6}{\includegraphics{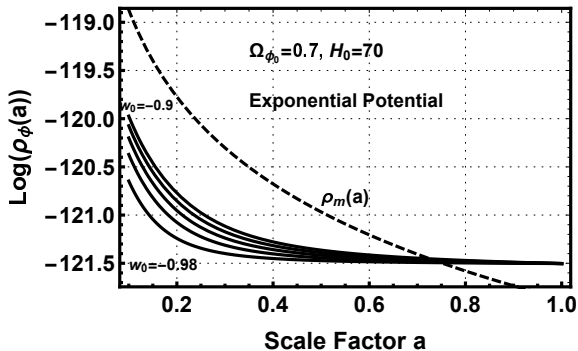}} 
\caption{The dark energy density values as a function of the scale factor for the exponential
potential.  As in fig~\ref{fig-lgrde} the dashed line shows the matter density.}
\label{fig-exrde}
\end{figure}

From the Einstein equation with mass 
\begin{equation} \label{eq-e1}
3H^2=\rho_m + \rho_{\phi}
\end{equation}
it is clear that 
\begin{equation} \label{eq-rphi}
\rho_{\phi} = 3 H^2 - \rho_m = 3H^2(a) -\frac{\rho_{m_0}}{a^3}
\end{equation}
for a flat universe.  Figures~\ref{fig-lgrde} and~\ref{fig-exrde} show the evolution
of the dark energy density for the logarithmic and exponential potentials respectively.
The dashed line in the figures shows the evolution of the matter density.  The reason
for the similarity of the quintessence evolution of $H(a)$ to the $\Lambda$CDM
evolution is shown in the figures.  The quintessence dark energy density evolves very
slowly in the current dark energy dominated epoch, mimicking the static cosmological
constant dark energy density.  The quintessence dark energy density only evolves 
significantly at high redshift in the matter dominated era. This is why the $H(a)$ evolution is 
essentially similar for the two cosmologies and may be true for most freezing cosmologies.  
There are thawing quintessence cosmologies \citep{sch08} however their potentials are 
extremely flat and must match the same value of $H_0$ as the freezing models. 

\section{The Dark Energy Equation of State} \label{s-deos}
\begin{figure}
\scalebox{.6}{\includegraphics{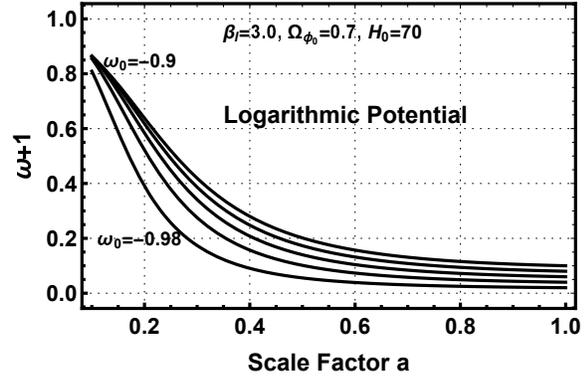}}
\caption{The evolution of $(w(a)+1)$ as a function of $a$ for the logarithmic dark energy
potential with $\beta_l=3$ and all five values of $w_0$. }
\label{fig-lgwpo}
\end{figure}

\begin{figure}
\scalebox{.6}{\includegraphics{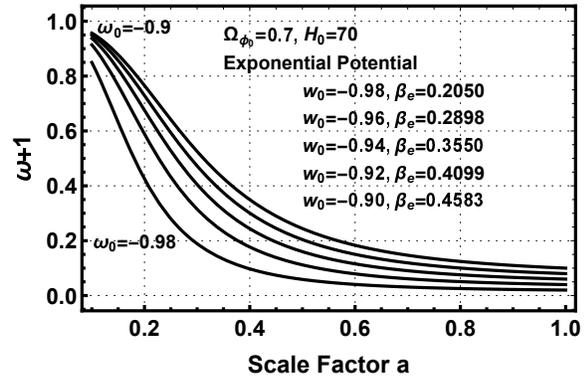}}
\caption{The evolution of $(w(a)+1)$ as a function of $a$ for the exponential dark energy
potential for the five $\beta_e$ values set by $(w_0+1) = 0.02, 0.04, 0.06, 0.08$ and 0.1. }
\label{fig-exwpo}
\end{figure}

A primary observational indicator of a dynamical cosmology is a dark energy equation of 
state different from the cosmological constant value of minus one.  From paper I
\begin{equation} \label{eq-wden}
1+w(\phi) = \frac{k^2\beta^2}{3}(1-\frac{4\rho_{m_0}a^{-3}}{3W^2})^{-1}
=\frac{k^2\beta^2}{3}(1-\Omega_m)^{-1}=\frac{k^2\beta^2(\phi)}{3\Omega_{\phi}}
\end{equation}
for a flat universe. Figure~\ref{fig-lgwpo} shows the evolution of $(w(a)+1)$ for the 
logarithmic dark energy potential with $\beta_l=3$ and all five values of $w_0$.
Figure~\ref{fig-exwpo} shows the evolution of $(w(a)+1)$ for the exponential dark energy
potential with the $\beta_e$ values set by $(w_0+1) = 0.02, 0.04, 0.06, 0.08$ and 0.1.

A common feature of all of the potentials in this paper and paper I is a very slow late time,
$a>0.5$ evolution of $w(a)$ with significant evolution for scale factors between 0.1 and
0.5.  This indicates that at least for the quintessence cosmology that high redshift observations
have the best chance of detecting the presence of dynamical dark energy.  The shapes of
the logarithmic and exponential potential $w(a)$ are quite similar, particularly for the 
lower values of $w_0$, while they are more divergent for the higher values.  Any determination
of the dark energy potential from the $w(a)$ tracks would require a secure knowledge of
$w_0$ and very accurate measurements of $w(a)$ at higher redshifts.  The required level
of accuracy is beyond current observational capabilities.  Detection of the predicted value of 
$w(a)\approx -0.5$ at $a=0.2$, $z=4$, however, might be possible with present techniques.
Further discussion of $w(a)$ observations occurs in sec.~\ref{ss-ds}.

\subsection{The Fundamental Constants} \label{ss-fc}
Paper I gives an extensive discussion of the evolution of the fundamental constants for both
the proton to electron mass ratio $\mu$ and the fine structure constant $\alpha$ in terms
of a change of $\phi$ and a coupling constant $\zeta_c$ where $c$ is $\mu$ or $\alpha$.
\begin{equation} \label{eq-dx}
\frac{\Delta c}{c} = \zeta_c k(\phi - \phi_0) = \zeta_c \int^a_1\beta(a')d \ln a', \hspace{0.5cm} c=\alpha,\mu
\end{equation}
\citep{nun04}.  The first equality is usually interpreted as the first term of a Taylor expansion 
of a possibly more complicated coupling.  The observational constraints on $\Delta \alpha /\alpha$ 
and $\Delta \mu / \mu$ are of the order $10^{-6}$ or less, justifying the assumption.
The last equality, not shown in paper I, explicitly shows the connection between the beta function 
and the evolution of the fundamental constants.  Sections~\ref{ss-sfl} and~\ref{ss-sfe} show the
transformation of $\beta(\phi)$ to $\beta(a)$ via the formulae for $\phi(a)$.

\begin{figure}
\scalebox{.6}{\includegraphics{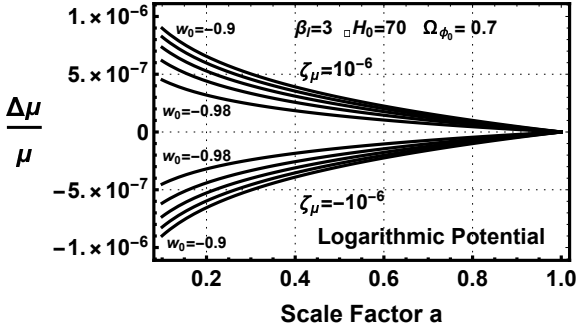}}  
\caption{The evolution of $\frac{\Delta \mu}{\mu}$ for the  logarithmic dark energy 
potential with $\beta_e = 3$ and  $w_0 = -0.98, -0.96,  -0.94, -0.92$ and -0.98.  The coupling 
constant $\zeta_{\mu}$ is set to $\pm 10^{-6}$ as an example.}
\label{fig-lgdmu}
\end{figure}

Figure~\ref{fig-lgdmu} shows the evolution of $\Delta \mu / \mu$ versus the scale factor for
the logarithmic potential with $\beta_l=3$ and the five values of $w_0$.  The positive and 
negative evolutions simply indicate that the coupling could have either a positive or negative
sign.  The coupling is arbitrarily set to $\pm 10^{-6}$ for the figure.  The evolution of the
fine structure constant is identical for the same coupling constant.  The evolution reflects 
the evolution of $\phi(a)$ since the coupling is assumed to be a constant.  As expected the higher
the deviation of $w_0$ is from minus one the larger the evolution of $\mu$.  Similar to the
power law potentials in paper I the evolution of $\Delta \mu / \mu$ relatively insensitive to
changes in $\beta_l$.

\begin{figure}
\scalebox{.6}{\includegraphics{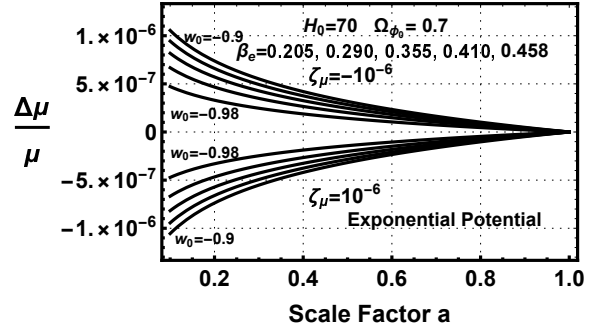}}  
\caption{The evolution of $\frac{\Delta \mu}{\mu}$ for the exponential dark energy potential
for the $\beta_e$ values set by $(w_0+1) = 0.02, 0.04, 0.06, 0.08$ and 0.1.  The coupling 
constant $\zeta_{\mu}$ is set to $\pm 10^{-6}$.}
\label{fig-expdmu}
\end{figure}

Figure~\ref{fig-expdmu} shows the evolution of $\mu$ for the exponential potential.  As 
described in section~\ref{ss-sfe} $w_0$ and $\beta_e$ are not independent variables in
the exponential case.  In fig.~\ref{fig-expdmu} the five values of $w_0$ are retained as in
fig.~\ref{fig-exphi} with the appropriate values of $\beta_e$ for each case.  The values of $\beta_e$
are shown in fig.~\ref{fig-expdmu}.  

\subsubsection{Observational Constraints on  $\frac{\Delta \mu}{\mu}$} \label{sss-muob}
\begin{figure}
\scalebox{.6}{\includegraphics{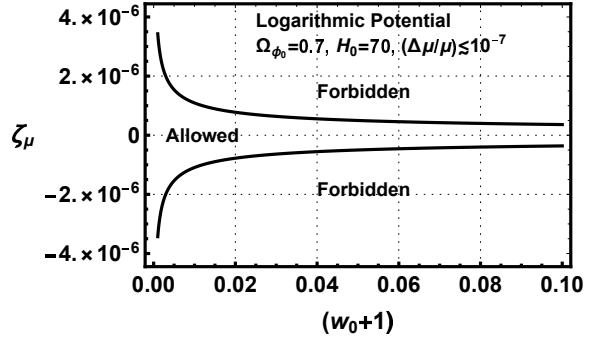}}  
\caption{The allowed and forbidden parameter spaces in the $\zeta_{\mu}$ - $w_0$ plane for
the logarithmic dark energy potential.}
\label{fig-lfc}
\end{figure}

\begin{figure}
\scalebox{.6}{\includegraphics{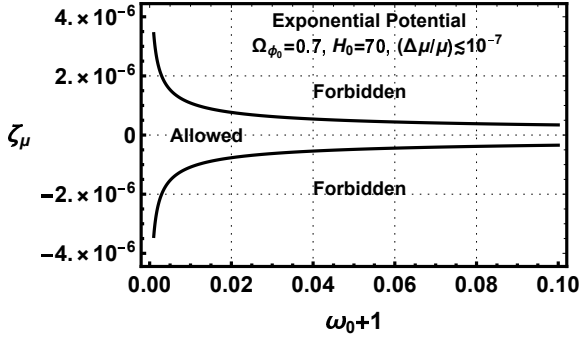}}  
\caption{The allowed and forbidden parameter spaces in the $\zeta_{\mu}$ - $w_0$ plane for
the exponential dark energy potential.}
\label{fig-efc}
\end{figure}

As discussed in paper I the primary constraint on a variation of $\mu$ is $\Delta \mu / \mu 
\le \pm 10^{-7}$ from
\citet{bag13} and \citet{kan15} at a redshift of 0.88582.  This measurement defines an allowed
and a forbidden parameter space in the $\zeta_{\mu}$ $w_0$ plane.  The first parameter,
$\zeta_{\mu}$, defines the limits on the allowed deviation from the standard model, $\zeta_{\mu}
= 0$, and the second, $w_0$, the allowed deviation from the cosmological constant, $(w_0+1)=0$.
The upper limit on $\zeta_{\mu}$ is given by
\begin{equation} \label{eq-uz}
\zeta_{\mu} = \frac{\Delta \mu / \mu}{\int_1^{a_{ob}}\beta(a')d\ln(a')}=\frac{\Delta \mu / \mu}
{\sqrt{3 \Omega_0(w_0+1)}\ln(a_{ob})}
\end{equation}
where $a_{ob}$ is the scale factor at the epoch of the observation.  The second equality shows
explicitly the dependence on $w_0$.  Figure~\ref{fig-lfc} shows the allowed and 
forbidden parameter space for the logarithmic dark energy potential 
and fig.~\ref{fig-efc} the parameter spaces for the exponential potential.
The plots start at $(w_0+1)=0.001$ to avoid the plus and minus infinite values of $\zeta_{\mu}$
at $(w_0+1)=0$.
The allowed parameter space contains the $\Lambda$CDM cosmology which is the 0,0 point in
the plots.  Although constrained to either small values of $\zeta_{\mu}$ or $(w_0+1)$ there
is still room in the allowed parameter space to accommodate quintessence.

\section{Relevant but not Directly Observable Parameters} \label{s-rp}
There are several cosmological parameters that are relevant but not directly observable.
Here two parameters, the time derivative of the scalar field and the dark energy pressure, 
are calculated as functions of the scale factor $a$.

\subsection{The Evolution of the Time Derivative of the Scalar} \label{ss-phidot}
\begin{figure}
\scalebox{.6}{\includegraphics{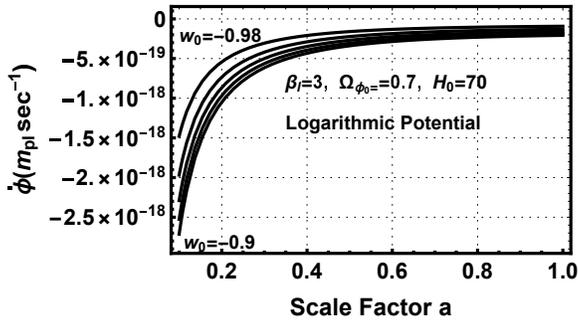}} 
\caption{The time derivative of the scalar for the logarithmic potential for the five values
of $w_0$ with $\beta_l$ held constant at three.}
\label{fig-lpdot}
\end{figure}

As shown in paper I the time derivative of the scalar $\dot{\phi}$ is simply the Hubble parameter 
times the beta function.
\begin{equation} \label{eq-pdot}
k\dot{\phi}= a\frac{k d\phi}{da}\frac{\dot{a}}{a}=\beta H 
\end{equation}
Figure~\ref{fig-lpdot} shows the evolution of $\dot{\phi}$ with respect to the scale factor 
$a$ for the logarithmic dark energy potential for the five values of $w_0$ with $\beta_l$
held constant at three.  The differences in the tracks is entirely due the differences in the 
beta function since the values of $H(a)$ are essentially invariant with respect to the input 
parameters as shown in section~\ref{ss-eh} and paper I.  The values of $\dot{\phi}$ are
negative because the logarithmic beta function is negative.

\begin{figure}
\scalebox{.6}{\includegraphics{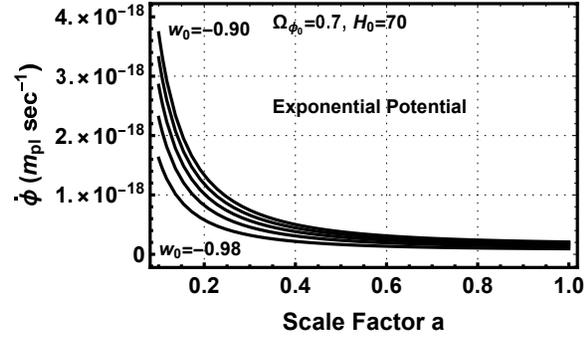}} 
\caption{The time derivative of the scalar for the exponential potential for the five values
of $w_0$ and their associated values of $\beta_e$.} 
\label{fig-epdot}
\end{figure}

Figure~\ref{fig-epdot} shows the tracks of $\dot{\phi}$ as a function of $a$ for
the exponential dark energy potential for the five values of $w_0$ and the $\beta_e$
values associated with them.
Both the logarithmic and the exponential have $\dot{\phi}$ values approaching zero at
the present time.

\subsection{The Evolution of the Dark Energy Pressure} \label{ss-dep}
The dark energy pressure comes from the second of the Einstein eqns.
$-2\dot{H}= \rho_m +\rho_{\phi}+p_{\phi}$ where $\dot{H} =-\frac{1}{2}
\dot{\phi}W_{,\phi}$. From eq~\ref{eq-difw}
\begin{equation} \label{eq-dwp}
W_{,\phi}=-\frac{2 \rho_m}{\beta W}-\frac{1}{2}\beta W
\end{equation}
which yields using $k\dot{\phi}=-\frac{\beta W}{2}$
\begin{equation} \label{eq-pp}
p_{\phi}=-2\rho_m-\frac{k^2}{2}\dot{\phi}^2+3H^2
\end{equation}

\begin{figure}
\scalebox{.6}{\includegraphics{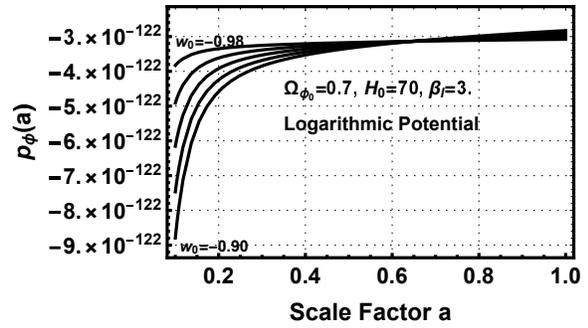}}  
\caption{The dark energy pressure for the five values of $w_0$ with a logarithmic
potential with $\beta_l=3$.}
\label{fig-lgdep}
\end{figure}

\begin{figure}
\scalebox{.6}{\includegraphics{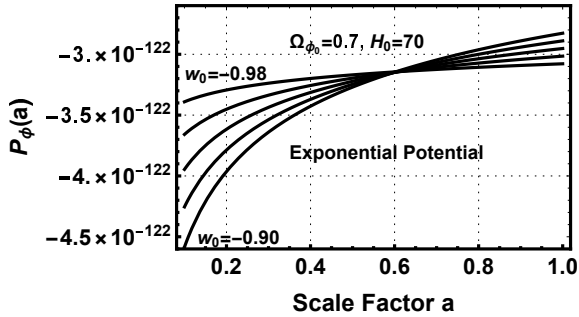}}  
\caption{The dark energy pressure for the five $\beta_e$, $w_0$ pairs with an exponential
potential.}
\label{fig-expdep}
\end{figure}

Figure~\ref{fig-lgdep} shows the evolution of the dark energy pressure for the logarithmic
dark energy potential for the five values of $w_0$ with $\beta_l=3.$ Since the pressure is 
negative the negative numbers rather than the logarithms are plotted. 
As expected from the dark energy density plots the $p_{\phi}$ tracks cross over themselves.
Figure~\ref{fig-expdep} shows the $p_{\phi}$ for the exponential potential for the five
$\beta_e$, $w_0$ pairs.

\section{Relevance of the Analysis} \label{s-sum}
This paper completes the investigation started in paper I of four common dark energy potentials 
in a quintessence cosmology.  Here the relevance of the findings to important cosmological and
new physics questions is examined.  The literature on determining cosmological parameters
based on observations is vast and it is not the purpose of this section to determine the veracity
of the various studies.  Instead the following points out which parameters calculated in this study
and paper I are relevant to the important questions and how they may differ from the current
body of work.

\subsection{Dynamical versus Static Dark Energy} \label{ss-ds}
What observations can discriminate between a dynamic dark energy quintessence cosmology
and a static dark energy $\Lambda$CDM universe?  An important finding is that due to
the flatness of the quintessence potentials in the dark energy dominated eras both cosmologies
predict essentially identical evolution of the Hubble parameter $H(a)$. $H(a)$ measurements, 
therefore, can not effectively discriminate between the two cases.  Measurements
that differed from the predicted evolution would, however, rule out both cosmologies.

\begin{figure}
\scalebox{.6}{\includegraphics{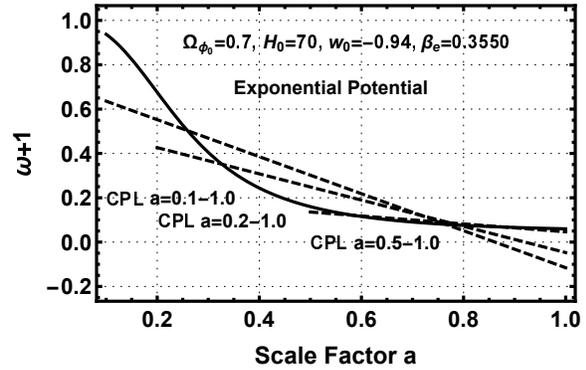}}  
\caption{Three CPL fits, dashed lines, to the evolution of $w(a)$ for an exponential 
potential with $w_0= -0.94$.  The fits to the full range and the range between $a=0.2$
and 1.0 produce false phantom crossings.}
\label{fig-cpl}
\end{figure}

Measurements of the dark energy equation of state $w(a)$ and the values of the fundamental
constants $\mu$ and $\alpha$ can discriminate between dynamical and static dark energy.
A confirmed observation of $w(a) \neq -1$ or a change in the value of a fundamental
constant would rule out $\Lambda$CDM but would be consistent with quintessence or
other dynamical dark energy cosmologies.  Tests for a value of $w(a) \neq -1$ eg. \citet{avs17}, 
are often conducted using the Chevallier-Polarsky-Linder (CPL) linear model  \citep{che01, lin03}.
\begin{equation} \label{eq-cpl}
w(a) = w_0+w_a(1-a)
\end{equation}
Examination of figs.~\ref{fig-lgwpo} and~\ref{fig-exwpo} indicates that the model is a 
reasonable fit at low redshifts but is a bad fit at high redshifts where $w(a)$ is evolving 
rapidly in the quintessence cosmology.  The tracks in these figures provide more realistic
templates to compare with observations than the CPL linear model.  The shapes of the 
$w(a)$ tracks suggest a possible reason why many observational studies seem to favor
phantom, $w<-1$, values of $w$ e.g. \citep{chen17}. Figure~\ref{fig-cpl} shows CPL 
fits to the $w_0=-0.94$ 
$w(a)$ evolution for an exponential potential over three scale factor ranges; the full
range between 0.1 and 1.0 ($z=9-0$), the range between 0.2 and 1.0 ($z=4-0$) and the
range between 0.5 and 1.0 ($z=1-0$).
\FloatBarrier
As expected the fit between $z=1$ and 0 is a good match but the two fits that include
the higher redshift evolution produce phantom values for $w_0$ in eqn.~\ref{eq-cpl}
even though the true evolution has no phantom values. It is also evident that observations 
at redshifts greater than one provide more leverage on constraining deviations of $w(a)$ 
from minus one than observations between redshifts one and zero.  

Measurements of the values of $\mu$ and $\alpha$ provide more precise constraints on
dynamical dark energy.  Figures~\ref{fig-lgdmu} and~\ref{fig-expdmu} show the expected
evolutionary tracks for $\mu$ with a coupling constant $\zeta_{\mu} = \pm 10^{-6}$ and
the five different values of $w_0$.  A single measurement, under the quintessence assumption
of homogeneous dark energy, determines the allowed parameter space for dynamical dark energy.
Figures.~\ref{fig-lfc} and ~\ref{fig-efc} show the allowed parameter space in the $(w_0+1)$,
$\zeta_{\mu}$ plane based on the observational constraint discussed in sec.~\ref{sss-muob}.  
Any point other than 0,0 in the plane requires dynamical dark energy, new physics or both.  

\subsection{The Dark Energy Potential} \label{ss-dep}
As with the question of dynamical versus static dark energy, the Hubble parameter yields
essentially no information on the functional form of the dark energy potential.  Although
not explicitly depicted here the tracks of the cosmological parameters, such as $w(a)$
for the logarithmic potential have the same insensitivity to the value of $\beta_l$ as 
shown for the power and inverse power law potential in paper I.  The $w(a)$ tracks
for the exponential potential, however, are sensitive to $\beta_e$ since the values of
$\beta_e$ and $w_0$ are coupled by eqn.~\ref{eq-bew}.

An accurate observational measurement of $w(a)$ at a particular scale factor or for a 
range of scale factors does not uniquely determine the dark energy potential.
Examination of figs.~\ref{fig-lgwpo} and~\ref{fig-exwpo} shows that for a given
coordinate in the $w(a)$,$a$ plane either a logarithmic or exponential potential
can match the coordinate by altering the value of $w_0$.  The tracks in the two
figures are for specific values of $w_0$ but all of the area between the minimal
and maximal tracks are covered by the range of $w_0$ between -0.9 and -0.98.
All of the area below the minimal -0.98 can be covered by making $w_0$
arbitrarily close to -1.0 and the area above the maximal -0.9 track can be covered
by making $w_0$ even further from -1.  The tracks in both figures have very
similar shapes, making it difficult to discriminate between the potentials even
with good knowledge of $w(a)$ over a large range of scale factors.  However,
if there is an accurate measurement of $w_0$ along with $w(a)$ at other scale
factors there is some leverage in determining the potential.  Of course any
determination of $w$ other than minus one at any epoch would be a 
significant finding.

\subsection{The Rate of Change of Fundamental Constants} \label{ss-crcfc}
\begin{figure}
\scalebox{.6}{\includegraphics{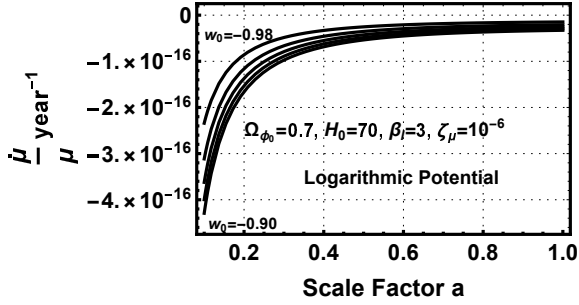}}  
\caption{The rate of change, $\dot{\mu}/\mu$ per year for the logarithmic potential
with a coupling constant of $\zeta_{\mu} = 10^{-6}$.}
\label{fig-lgdmr}
\end{figure}

\begin{figure}
\scalebox{.6}{\includegraphics{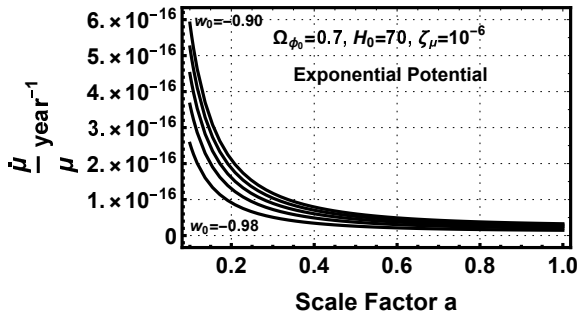}}  
\caption{The rate of change, $\dot{\mu}/\mu$ per year for the exponential potential
with a coupling constant of $\zeta_{\mu} = 10^{-6}$.}
\label{fig-exdmr}
\end{figure}

Laboratory constraints on the rate of change of fundamental constants is another
check on the possibility of dynamical dark energy.  Figures~\ref{fig-lgdmu}
and~\ref{fig-expdmu} indicate that the rate of change of $\mu$ and $\alpha$ in a
quintessence freezing cosmology is slowing down in the current epoch.  
Figures~\ref{fig-lgdmr} and~\ref{fig-exdmr}
show the rate of change, $\dot{\mu}/\mu$ per year for the logarithmic and exponential
dark energy potentials with a coupling constant of $\zeta_{\mu}= 10^{-6}$.
The proton to electron ratio is used in the example but the fine structure constant $\alpha$ 
has exactly the same track if its coupling constant is also $\zeta_{\alpha} = 10^{-6}$.
\begin{table}
\begin{center}
\begin{tabular}{|c|c|c|c|c|}
    \hline
       & \multicolumn{4}{c}{$\dot{\mu}/\mu$ in $10^{-17} m_{pl}$ per year} \\ 
   \hline
   & \multicolumn{2}{|c|}{Logarithmic} &  \multicolumn{2}{|c|}{Exponential}\\
   \hline
$w_0$&a=0.1&a=1.0&a=0.1&a=1.0\\
  \hline
-0.98 & -23.4 & -1.47 & 25.6 & 1.47 \\
-0.96 & -31.1 & -2.07 & 36.5 & 2.07 \\
-0.94 & -36.2 & -2.54 & 45.0 & 2.54 \\
-0.92 & -40.0 & -2.93 & 52.4 & 2.93 \\
-0.90 & -42.9 & -3.28 & 59.0 & 3.28 \\
    \hline
\end{tabular}
\end{center}
\caption{$\dot{\mu}/\mu$ per year for the logarithmic and exponential
dark energy potentials at scale factors of 0.1 and 1.0 for the five values
of the current dark energy equation of state $w_0$ and a coupling constant of
$\pm 10^{-6}$} \label{tab-rfc}
\end{table}
It is clear from fig.~\ref{fig-lgdmr} and~\ref{fig-exdmr} that in a quintessence
freezing cosmology the current rate of change of the fundamental constants is 
significantly less than rate at high redshift.  Table~\ref{tab-rfc} shows the rate
of change in units of $10^{-17} m_{pl}$ per year for $\mu$ at scale factors of 
0.1 and 1.0 for the logarithmic and exponential potentials for the five values of 
$w_0$ and a coupling constant of $+10^{-6}$.  The signs between the two potentials 
are opposite and would be reversed for a negative coupling constant. 
The current rates of change are essentially the 
same between the two potentials but diverge at a scale factor of 0.1. The average 
current rate of change is roughly 18 times less than the rate of change at a scale 
factor of 0.1.  Current laboratory bounds~\citep{god14} are $\dot{\mu}/\mu =
(0.2 \pm 1.1) 10^{-16}$ year$^{-1}$ and $\dot{\alpha}/\alpha =(-0.7 \pm 2.1)
10^{-17}$ year$^{-1}$.  Matching the cosmological observational bounds on
$\Delta \mu / \mu$ discussed in sec.~\ref{sss-muob} with a coupling constant of
$\pm 10^{-6}$ requires $(w_0+1) \leq 0.02$ which is the first row in 
table~\ref{tab-rfc}. This sets a limit a factor of ten below the laboratory limit.
Unlike the laboratory limits the cosmological limit on $\Delta \mu / \mu$ is
more stringent than the limit on $\Delta \alpha / \alpha$.

\subsection{Checking on the Swampland} \label{ss-cs}
String theory postulates a vast landscape of vacua that is surrounded by an
even more vast landscape, termed the swampland, of consistent looking
scalar field theories that are inconsistent with a quantum field theory of gravity
\citep{vaf05,agr18}.  Put another way the swampland is the landscape of valid 
scalar field theories that are incompatible with quantum gravity \citep{hei18}.
Given the current interest in the swampland it is worthwhile to determine whether
quintessence with the potentials considered here and in paper I dwells in the 
swampland.  The boundaries of the swampland are usually defined by two
conjectures.  The first conjecture is that the change in the scalar should be
$\Delta \phi < \sim O(1)$ and the second is that $\Delta V/V \geq \sim O(1)$.  If
either of these conjectures are violated then the cosmology is in the swampland.
It is not entirely clear how restrictive of order 1 is or exactly what range of scale
factors $\Delta \phi$ and $\Delta V$  encompass.  It is obvious that $\Lambda$CDM
is in the swampland since $\Delta V = 0$.

The quintessence models considered here and in paper I certainly live near the
swampland with perhaps one foot in the swamp and one foot dry depending on
how of order 1 is interpreted.  The swampland parameters for the potentials
in this paper and paper I are shown in Table~\ref{tab-swamp}.  Both $\Delta$
values are for the scale factor range between 0.1 and 1.0.  The potential $V$
in $\Delta V/V$ is the current day potential.  The X in the inverse power law
parameters for $w_0=-0.90$ indicate that this is not a valid solution as shown
in paper I.
\begin{table}
\begin{center}
\begin{tabular}{ccccccccc}
    \hline
       & \multicolumn{8}{c}{Swampland Parameters} \\ 
   \hline
   & \multicolumn{2}{|c|}{Log} & \multicolumn{2}{|c|}{Exp}&  \multicolumn{2}{|c|}{Pow}&  \multicolumn{2}{|c|}{Inv Pow}\\
   \hline
$w_0$&$\Delta \phi$&$\frac{\Delta V}{V}$&$\Delta \phi$&$\frac{\Delta V}{V}$&$\Delta \phi$&$\frac{\Delta V}{V}$&$\Delta \phi$&$\frac{\Delta V}{V}$\\
  \hline
-0.98 \vline & 0.45 & 0.09 \vline & -0.47 & 0.10 \vline &0.55  &0.10 \vline &-0.58 &0.11 \\
-0.96 \vline & 0.62 & 0.18 \vline & -0.67 & 0.21 \vline &0.76  &0.20 \vline &-0.84 &0.23  \\
-0.94 \vline & 0.73 & 0.26 \vline & -0.82 & 0.34 \vline &0.92  &0.30 \vline &-1.06 &0.38  \\
-0.92 \vline & 0.82 & 0.34 \vline & -0.94 & 0.47 \vline &1.04  &0.41 \vline &-1.26 &0.56  \\
-0.90 \vline & 0.90 & 0.42 \vline & -1.06 & 0.62 \vline &1.14  &0.52 \vline &X &X  \\
    \hline
\end{tabular}
\end{center}
\caption{The two swampland parameters for the potentials in this paper and paper I.
The units of $\Delta \phi$ are Planck masses.  The $\beta_{l,p,i}$ values are 3 for all
potentials except for the exponential potential which uses the $\beta_e$ value appropriate
to the $w_0$ value.}
\label{tab-swamp}
\end{table}

All of the exponential and logarithmic potential cases considered here satisfy the condition on 
$\Delta \phi$ under the assumption that -1.06 is of order 1.  The power and inverse power
law $\Delta \phi$ entries for the three values of $w_0$ closest to minus one, the most likely 
values, also satisfy the $\Delta \phi$ conjecture, the dry foot. None of the $\Delta V/V$ entries 
strictly satisfy the associated conjecture, the wet foot.  Very recent work by \citet{kin18}
suggests that this is a feature common to most single scalar field cosmologies.  Since the 
potentials $V(\phi)$ are functions of the scalar $\phi$ larger values of $\Delta V$ require 
larger changes in $\phi$ which, as table~\ref{tab-swamp} shows requires larger deviations 
of $w$ from minus one and drives the $\Delta \phi$ values higher which could result in violating
the $\Delta \phi$ conjecture. \citet{obi18} have also suggested
a criterion that  $|\phi|<1$ in Planck units which is not satisfied by the scalars in this work.
It is not the purpose of this discussion to determine whether having one foot in the swamp
is a good or bad thing but rather to simply show where quintessence with the potentials examined
here lies with respect to the swampland boundaries.

\section{Conclusions} \label{s-con}
This and paper I show that the beta function formalism provides an effective way to calculate
accurate solutions for cosmological parameters as a function of the scale factor $a$.  For the
most part the solutions are analytic functions utilizing known mathematical functions.  The 
superpotential for the logarithmic dark energy potential, however, required an easily calculated 
numerical integral.  The two papers also demonstrate the application of the beta function
formalism and can act as a guide to the extension of the formalism to other potentials and
cosmologies.

\label{lastpage}
\end{document}